\documentclass[usenatbib]{mn2e} % Normal MNRAS format
\usepackage{longtable} % Allows table to span multiple pages
\usepackage{amssymb, textcomp, verbatim, graphicx, times, multirow} % hyperref, 

\def\apj{ApJ}% % Astrophysical Journal
\def\mnras{MNRAS}% % Monthly Notices of the RAS
\def\sun{$_\odot$}
\def\deg{$^o$}

\title[SS Cyg Feb 2016 Anomalous Outburst]{Rapid Radio Flaring during an Anomalous Outburst of SS Cyg}
\author[Mooley et al.]{K. P. Mooley $^{1,9}$, J. C. A. Miller-Jones$^2$,  R. P. Fender$^1$, G. R. Sivakoff$^3$, C. Rumsey$^4$, \and
Y. Perrott$^4$, D. Titterington$^4$, K. Grainge$^5$, T. D. Russell$^{2,6}$, S. H. Carey$^4$, J. Hickish$^4$, \and 
N. Razavi-Ghods$^4$, A. Scaife$^5$, P. Scott$^4$, E. O. Waagen$^7$
\\
  $^1$ Centre for Astrophysical Surveys, University of Oxford, Denys Wilkinson Building, Keble Road, Oxford OX1 3RH\\
  $^2$ International Centre for Radio Astronomy Research - Curtin University, GPO Box U1987, Perth, WA 6845, Australia\\
  $^3$ Department of Physics, University of Alberta, 4-181 CCIS, Edmonton, AB T6G 2E1, Canada\\
  $^4$ Astrophysics Group, Cavendish Laboratory, 19 J. J. Thomson Avenue, Cambridge CB3 0HE, UK\\
  $^5$ University of Manchester, Alan Turing Building, Oxford Road, Manchester M13 9PL, UK\\
  $^6$ Anton Pannekoek Institute for Astronomy, University of Amsterdam, PO Box 94249, NL-1090 GE Amsterdam, Netherlands\\
  $^7$ American Association of Variable Star Observers, 49 Bay State Road, Cambridge, MA 02138, USA \\
  $^9$ Hintze Research Fellow; Email: kunal.mooley@physics.ox.ac.uk}

\begin{document}
\maketitle

% We report the discovery of rapid radio flaring in the dwarf nova, SS Cyg, together with the detection of a fast luminous flare towards the end of the outburst. 
% Our results provide new insights into the connection between accretion and jet production in accreting white dwarf systems. 
% These results will have a significant impact on the development of research in the field of dwarf novae, and will benefit from rapid publication.
% Our manuscript abides by the norms for a Letter (5 pages; note that Tables 1 and 2 are online-only). 

%--------------------------------------------------------------------------------------------
% ABSTRACT AND KEYWORDS
%--------------------------------------------------------------------------------------------
\begin{abstract}
The connection between accretion and jet production in accreting white dwarf binary systems, especially dwarf novae, is not well understood.
Radio wavelengths provide key insights into the mechanisms responsible for accelerating electrons, including jets and outflows.
% , and into the energetics associated with these processes.
Here we present densely-sampled radio coverage, obtained with the Arcminute MicroKelvin Imager Large Array, of the dwarf nova SS Cyg during its February 2016 anomalous outburst.
The outburst displayed a slower rise (3 days mag$^{-1}$) in the optical than typical ones, and lasted for more than 3 weeks.
Rapid radio flaring on timescales $<$1 hour was seen throughout the outburst.
The most intriguing behavior in the radio was towards the end of the outburst where a fast, luminous (``giant''), flare peaking at $\sim$20 mJy and lasting for 15 minutes was observed.
This is the first time that such a flare has been observed in SS Cyg, and insufficient coverage could explain its non-detection in previous outbursts.
These data, together with past radio observations, are consistent with synchrotron emission from plasma ejection events as being the origin of the radio flares.
However, the production of the giant flare during the declining accretion rate phase remains unexplained within the standard accretion-jet framework and 
appears to be markedly different to similar patterns of behavior in X-ray binaries.
\end{abstract}


%--------------------------------------------------------------------------------------------
% SECTION: INTRODUCTION
%--------------------------------------------------------------------------------------------
\section{Introduction}\label{sec:intro}
\vspace{-0.1in}
Dwarf novae (DNe) are binary systems containing a non-magnetic ($B_{\rm surface}<10^6$ G) white dwarf actively accreting from a Roche lobe-filling main-sequence companion via an accretion disk \citep{warner1995}.
All known DNe undergo episodic outbursts, which last between a few days and a few years, recur between 10 days and several decades, and result in brightening by 2--8 magnitudes in the optical \citep[e.g.][]{coppejans2016a}.
The disk instability model, in which the accretion disk cycles between cool quiescent states and hot outburst states, provides a consistent framework for explaining these outbursts \citep[e.g.][]{osaki1974,meyer1981,lasota2001}.
DN outbursts seem to be analogous to X-ray binary (XRB) outbursts \cite[e.g.][]{kuulkers1996}, and the similarities between their hardness-intensity diagrams (disk-fraction luminosity diagram, or DFLD, in the case of DNe) 
together with the presence of transient radio emission suggests that the connection between the outburst phase and the radio (jet) emission could be similar as well \citep{kording2008}.

While some classes of accreting white dwarf systems, such as symbiotic stars, and magnetic cataclysmic variables, were detected in the 
radio\footnote{Radio emission is also expected from supersoft X-ray sources, which show evidence for jets through pairs of Doppler shifted emission lines.},
DNe were, until recently, thought to be weak radio emitters \citep[e.g.][]{soker2004}.
Even for the few DNe that were detected in the radio during outbursts, the detection was not reproducible \citep[e.g.][]{benz1983,benz1996}. 
\cite{coppejans2015} and \cite{coppejans2016b} showed that many of the non-magnetic accreting white dwarf systems are faint ($\lesssim$0.1 mJy) and variable on $\sim$hours timescales in the radio, 
and that the sensitivities, time-resolutions or timing of previous observations were insufficient to detect these fainter systems.
DNe have now been shown to be repeating radio emitters during outbursts \citep{coppejans2016b,russell2016}.
The brightness temperature from VLBI imaging, the outburst track in the DFLD, the radio spectral indices, and the shape of the radio light curves have been used to 
argue in support of synchrotron emission from a transient jet as the cause of the radio emission during DNe outbursts \citep{kording2008,russell2016,coppejans2016b}.

The quintessential DN, SS Cyg, is one of the best studied systems of its class across the electromagnetic spectrum.
This system consists of a $\sim$1 M\sun white dwarf with a K5V companion, with the inclination of the binary being close to 45\deg \citep{north2002,kording2008}.
The precise VLBI parallax distance measurement\footnote{The Gaia distance is $117\pm4$ pc \citep{gaia2016}.} of $114\pm2$ pc \citep{miller-jones2013} allows the study of the energetics of this system at a 
level of precision impossible to achieve for most accreting systems.
Long-term optical monitoring of SS Cyg has revealed regular outbursts lasting $\sim$5--20 days and recurring on timescales between $\sim$20--60 days \citep{cannizzo1992}.
The majority of the outbursts have rise and decay rates (in the optical) that are remarkably similar from one outburst to another, and scattered around 0.5 days mag$^{-1}$ and 2.5 days mag$^{-1}$ respectively.
About 10\% of the outbursts, referred to as ``anomalous'', exhibit a slow rise of $\gtrsim$1.5 days mag$^{-1}$ in the optical \citep{cannizzo1998}.
X-ray and ultraviolet observations suggest that accretion in SS Cyg (and similarly in other dwarf novae) occurs via a ``boundary region'' between the inner accretion disk and the surface of the white dwarf.
The boundary region becomes optically thick at X-ray frequencies at the onset of an outburst and may play a role in jet production in SS Cyg \citep{wheatley2003,russell2016}.
Analogous to XRBs, SS Cyg is expected to have a steady jet during the rise phase of an outburst and discrete plasma ejection after the subsequent spectral softening \citep{kording2008,miller-jones2011}.

% Briefly describe the X-ray, UV and radio data taken so far and a sentence about their interpretation.
% Similar HID cycle => jets should be present \citep{kording2008,miller-jones2011}.
% 
% \citep{kording2008,russell2016}.
% Talk about the rise and decay rates of the optical light curves during normal and anomalous outbursts (Cannizzo \& Mattei 1998)?
% Radio, X-ray, UV and optical data - what do they tell us so far - boundary layer, optically thick, etc.
% 

In February 2016, SS Cyg underwent an anomalous outburst, which lasted for about 3 weeks (at V$>$11 mag).
The outburst showed a slow rise of about 3 days mag$^{-1}$ in the optical for one week and then transformed into a standard broad outburst before reaching the peak.
Here, we report on high-cadence radio observations of this anomalous outburst of SS Cyg.
In \S\ref{sec:obs} we describe the optical, radio and X-ray observations, and present our analysis and discussion in \S\ref{sec:discussion}.

%--------------------------------------------------------------------------------------------
% SECTION: OBSERVATIONS AND DATA
%--------------------------------------------------------------------------------------------
\vspace{-0.2in}
\section{Observations and Data Processing}\label{sec:obs}

\subsection{Optical}
We requested close monitoring of SS Cyg and immediate submission of observations to the American Association of Variable Star 
Observers (AAVSO; Special Notice \#412, Alert Notice 536) in order to catch the rise phase of the outburst and obtain a well-sampled optical light curve.
Good sampling at the beginning of the outburst was necessary for triggering the radio observations (see below).
The data were downloaded from the AAVSO website\footnote{https://www.aavso.org/data-download}.
We used the V band magnitudes for our analysis.
The optical light curve is shown in Figure~\ref{fig:main_lc}.

\vspace{-0.2in}
\subsection{Radio}
We monitored the AAVSO light curve, and once the optical light curve reached 11.5 mag in the pre-validated V band (on 2016 Feb 11), 
we triggered the Arcminute MicroKelvin Imager Large Array \citep[AMI-LA;][]{zwart2008} radio telescope.
Observations were made with the new digital correlator having 4096 channels across a 5 GHz bandwidth between 13--18 GHz.
SS Cyg was monitored for about 10 hours every day throughout the $\sim$3 weeks of outburst.
The phase calibrator, J2153+4322, was observed every 12 minutes for about 1.5 minutes.
The log of AMI-LA observations is given in Table~\ref{tab:observations}.

The AMI-LA data were binned to 8$\times$0.625 GHz channels and processed (RFI excision and calibration) with a fully-automated pipeline, AMI-REDUCE \citep[e.g.][]{davies2009,perrott2013}.
Daily measurements of 3C48 and 3C286 were used for the absolute flux calibration, which is good to about 10\%.
The calibrated amplitudes for J2153+4322 showed variability of 20\% (peak-to-peak) around the mean of 250 mJy at 15.5 GHz within each 10-hour observing block.
This variability in the complex gain calibrator is likely a combination of unmodeled gain variations and pointing errors\footnote{We note, however, that significant work has gone into improving 
the AMI-LA amplitude gain calibration and empirical pointing model.}.
Hence we recalibrated the data using fixed values for the flux density ($S_{15.5GHz} = 260$ mJy) and spectral index ($\alpha=-0.15$; $S \propto \nu^\alpha$) of J2153+4322, which are our best estimates based on 
our knowledge of the instrument and the data.
Conservative uncertainties in these estimates are 5\% (flux density) and 0.1 (spectral index), and hence we conclude that the error in flux density is dominated by the absolute flux scale uncertainty.
While there are no published 15 GHz flux density measurements for J2153+4322, the flux density and the spectral index that we have used are consistent with its 8.4 GHz flux densities 
measured in 2011\footnote{$\sim$280 mJy from http://astrogeo.org/v2m/maps/bc196zc/ and \cite{russell2016}, for example}.

The calibrated and RFI-flagged data were then imported into CASA. 
Since the AMI-LA synthesized beam is approximately 30\arcsec, the $\sim$1 mJy radio source 22.3\arcsec~ to the North-West of SS Cyg causes confusion and had to be subtracted from the UV plane.
We used the interactive mode of CASA {\tt clean} to define a restricted clean box at the coordinates of the confusing source (the clean box does not overlap with the location of SS Cyg) 
and derived a model for each AMI-LA observation.
We then loaded and subtracted the model from the CASA measurement sets using the tasks {\tt ft} and {\tt uvsub}.
We then split each 10-hour observation into shorter intervals as needed, according to the variability seen in the source, and imaged each interval using CASA {\tt clean}.
The flux density of SS Cyg was measured in the resulting 512$\times$512 pix$^2$ (4\arcsec pix$^{-1}$) images using the {\tt pyfits} module.

\begin{figure*}
\centering
\includegraphics[width=5.1in]{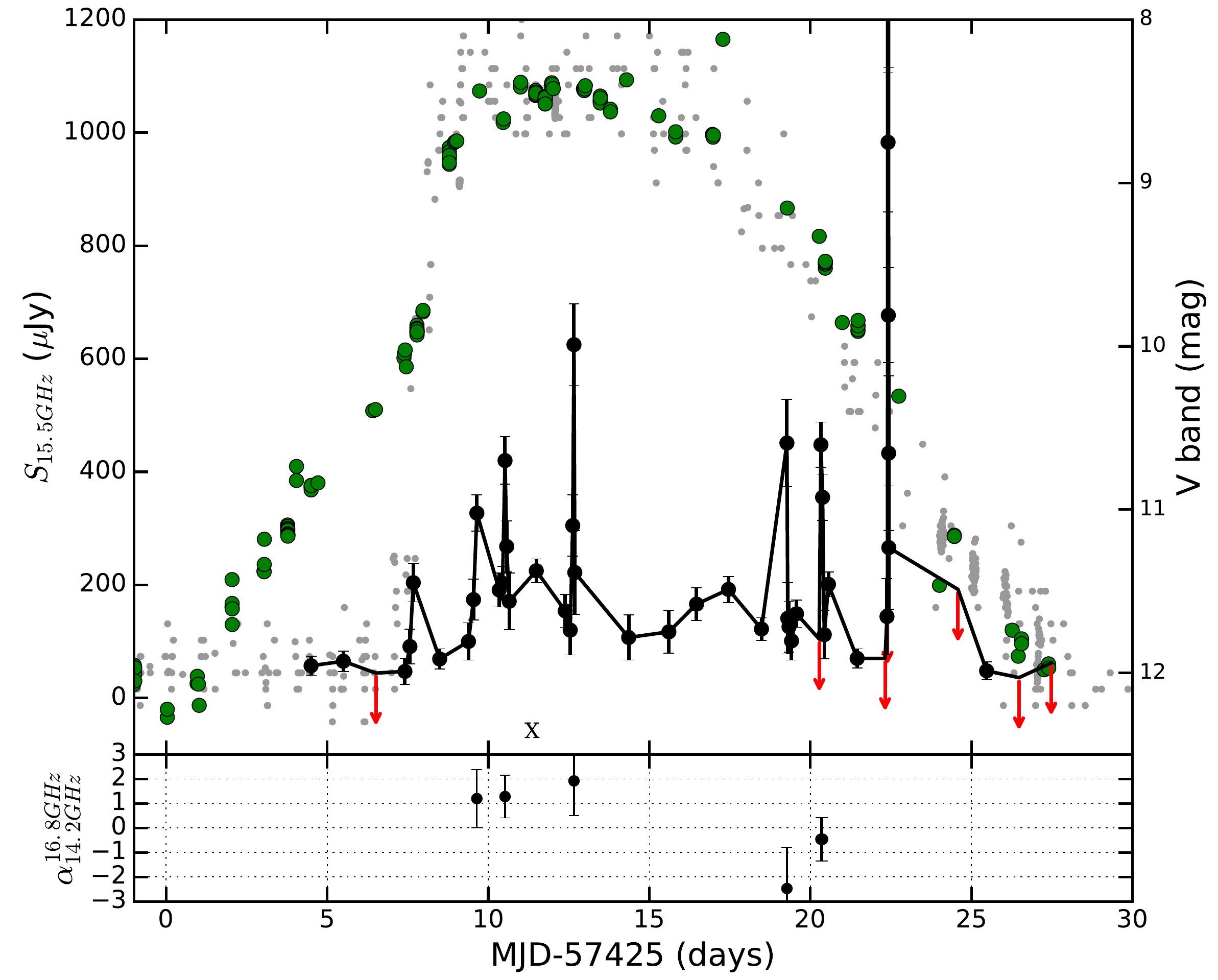}
\caption{\it The 15.5 GHz AMI-LA (black) and V-band AAVSO (green) light curves of SS Cyg during the February 2016 anomalous outburst.
% The radio measurements are at 15.5 GHz from the AMI-LA telescope and the optical data is V band from AAVSO.
Upper limits (2$\sigma$) in the radio are shown as red arrows.
For comparison, the optical light curve for a normal outburst presented by \citet{kording2008} is shown in grey (aligned at the steep rise).
The ``X'' symbol denotes the time of the X-ray observation (see \ref{sec:obs:xray} for details).
The ``giant'' flare peaking at $\sim$18 mJy, seen on MJD 57447 (29 February 2016) and lasting for $\sim$10 minutes, is shown in more detail in Figure~\ref{fig:bigflare_lc}. 
The upper limits on the day of the giant flare lie outside the time interval shown in Figure~\ref{fig:bigflare_lc}. 
Note that the peak of the giant flare extends much beyond the scale of this figure.
The bottom panel shows the spectral indices ($S \propto \nu^\alpha$; between 14.25 GHz and 16.75 GHz) during peaks of flaring events seen at 15.5 GHz.
}
\label{fig:main_lc}
\end{figure*}

The AMI-LA light curve is shown in Figure~\ref{fig:main_lc}.
We present a discussion of it in \S\ref{sec:discussion}.

\vspace{-0.3in}
\subsection{X-ray}\label{sec:obs:xray}

A 1 ks Swift XRT observation\footnote{Some of the requested XRT observations were interrupted by GRB observations.} was carried out on 2016 Feb 18.136 (start time; MJD 57436.136).
In 971.7 s of data, we detect a count rate of 1.887$\pm$0.049 per second in the 0.3--10 keV band.
The X-ray observation was not strictly simultaneous with any optical or radio observations.

A multi-temperature plasma emission model is often used to describe the X-ray spectra of dwarf novae, in which the boundary layer is heated to $\sim$10$^8$ K during outburst \citep[e.g.][]{wheatley2003,pandel2005}.
% The cemekl model is often used to describe the X-ray spectra of dwarf novae, which can reach as high as 10$^8$ K.
Accordingly, we fit a cemekl model \citep[the emission measures in this model follow a power law in temperature up to a certain maximum;][]{done1997} to the binned XRT spectrum using {\tt XSpec} and obtained a maximum temperature of $19.6^{+7.0}_{-4.3}$ keV.
% Note that in cemekl, a grid of MEKAL models are co-added, with the emission measures following a with index $\alpha$ and up to kT$_{\rm max}$.
% , which is qualitatively similar to model plT in Table 3 of Done \& Osborne 1997.
% Since we only care about this for a flux, I think this is ok. (Note this spectrum's highest valid bin cutsoff before the iron line at 6.4 eV, so no need for the "g" component in the Done & Osborne paper)
In the 0.5--10 keV, 1--10 keV, and 2--10 keV bands, this model has a flux of 6.5, 5.5, and 4.3 ($\times 10^{11}$) erg s$^{-1}$ cm$^{-2}$ respectively.
At a distance of 114 pc, we get a luminosity of $L_X = 1.7 \times 10^{31}$ erg s$^{-1}$ in the 1--10 keV band.

%-----------------------------Figure Start------------------------------
%-----------------------------Figure End--------------------------------

%-----------------------------Figure Start------------------------------
\begin{figure*}
\centering
\includegraphics[width=5.1in]{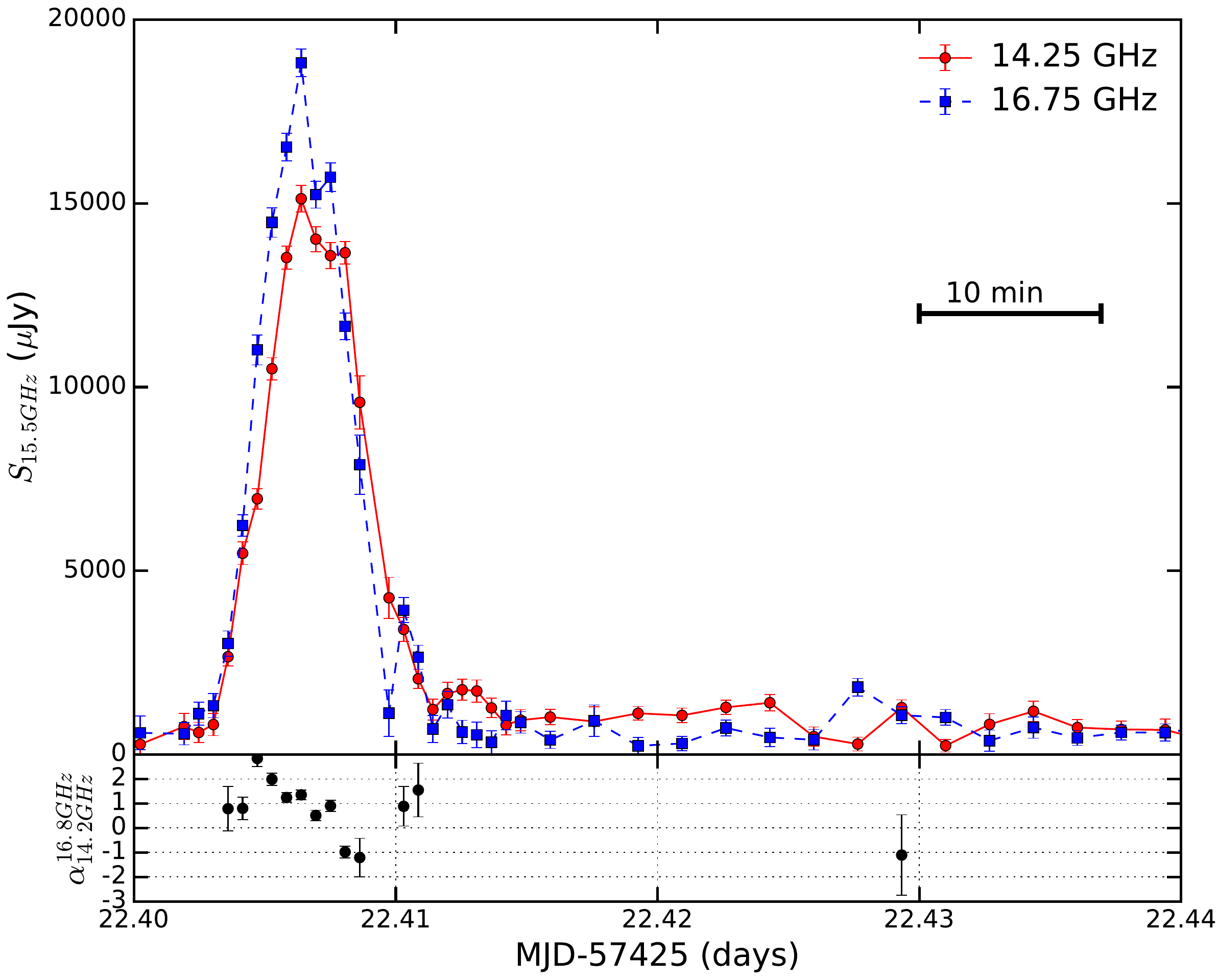}
\caption{\it The radio light curves at 14.25 GHz and 16.75 GHz during the SS Cyg ``giant'' flare on MJD 57447 (29 February), using data from Table~\ref{tab:observations_giantflare}.
The bottom panel shows the spectral indices between these two frequencies.
None of the optical observations were strictly simultaneous with this flare.}
\label{fig:bigflare_lc}
\end{figure*}
%-----------------------------Figure End--------------------------------

%--------------------------------------------------------------------------------------------
% SECTION: DISCUSSION
%--------------------------------------------------------------------------------------------
\vspace{-0.2in}
\section{Results \& Discussion}\label{sec:discussion}

% What did we see? 
% - When did the radio emission begin to rise
% - Fast flares, timescales ~30 min
% - What does our X-ray observation suggest?
% - core is switched on at all times, or does it switch off?
% - brightness temperature implied by the big flare: 
% - activity switches off and then the massive flare
% - there is structure in the big flare
% 
% - probably an ejection event
% - unexplained, brightest 

Our AMI-LA observations of SS Cyg during the February 2016 anomalous outburst represent the most densely-sampled radio coverage of any dwarf nova thus far.
The radio light curve at 15.5 GHz is shown in Figure~\ref{fig:main_lc} along with the optical light curve.
Our radio coverage began as soon as the optical light curve reached $V=11$ mag. (on 11 February 2016), and we find evidence for radio emission from 
SS Cyg right from our first observation, at a 15.5 GHz flux density of 57$\pm$17 $\mu$Jy.
Our previous AMI-LA observation from 02 February gave a non-detection, with a 2$\sigma$ upper limit\footnote{Although the convention is to quote 3$\sigma$ upper limits, 
the known coordinates of SS Cyg together with our knowledge of the noise properties and the manual inspection of radio images allow us to confidently place 
the detection threshold at 2$\sigma$. $<$1 spurious detection is expected at this level.}
% However, note that the RMS noise values that we have used do not include the absolute flux scale uncertainty.} 
of 86 $\mu$Jy.
We do not detect any significant intra-day and inter-day variability in the observations carried out between 11--13 February (MJD 57429--57431).
% after the onset of the slow rise in the optical light curve.

We detected significant variability on $\sim$hour timescales starting on 14 February.
A comparison with a ``normal'' optical outburst (grey points in Figure~\ref{fig:main_lc}) suggests that the flaring started when the normal optical outburst would have been expected to begin.
Further observations are necessary to verify whether this is indeed the case during a normal optical outburst.
The first flare occurred towards the end of our AMI-LA observation carried out on 14 February (MJD 57432.6726), at a peak flux density of $204\pm34$ $\mu$Jy at 15.5 GHz.
Subsequently, a series of rapid flares, with rise times between 5--30 minutes were seen throughout the outburst.
While the majority of the radio flares peaked at sub-mJy flux densities, we detected a very luminous flare towards the end of the outburst, peaking at $18.0\pm0.4$ mJy at 15.5 GHz 
and lasting for only $\sim$15 minutes.
The radio light curves at 14.25 GHz and 16.75 GHz\footnote{Our flux density values at 15.5 GHz are the mean over the whole 13--18 GHz band, while the 14.25 GHz and 16.75 GHz measurements 
are means over the 13--15.5 GHz and 15.5--18 GHz bands.} for this fast luminous flare (``giant'' flare) are shown in Figure~\ref{fig:bigflare_lc}.
There is some evidence of radio brightening of SS Cyg towards the end of the April 2007 and April 2016 outbursts as well (K\"ording et al. in prep, Miller-Jones et al. in prep), which 
may suggest the presence of such a giant flare is a common feature of outbursts in SS Cyg.
If so, then such a flare could have been missed in the majority of the past radio observations \citep[those reported by][]{russell2016} due to their much sparser time coverage.
Overall, during the outburst, the radio emission lasted for 3 weeks (11 February to 02 March 2016, i.e. MJD 57432--57449; above our 2$\sigma$ detection threshold of $\sim$40 $\mu$Jy).

% Give a general description of the flares
The initial radio flares, seen over a span of $\sim$1 week (between MJD 57432 and 57439), had rise times of tens of minutes and at peak they had an optically thick spectrum at 15.5 GHz.
Each successive flare was seen to have a peak flux density higher than the previous one.
If this flaring activity is related to plasma ejection events, then the optically thick spectrum (with $0\lesssim\alpha_{14.2GHz}^{16.8GHz}\lesssim2$) 
suggests that the particle acceleration time is shorter than the expansion timescale of the ejected blobs.
% to synchrotron self absorption.
This initial set of flares, which continued into the peak/flattening of the optical light curve, was followed by a few days where the radio emission was fairly steady or rising gradually.
The 15.5 GHz flux density was between 100--200 $\mu$Jy during this phase.
This steady period may be related to the ``plateau'' phase seen during past radio observations of SS Cyg \citep[e.g.][]{kording2008,russell2016}, although the coverage of these previous observations 
was very sparse.
Between MJD 57443 and 57446, the flaring activity resumed and during this period the overall spectrum at the flare peak was optically thinner than the earlier flares.
In general, the flare peaks are short lived and hard to define, but probably represent optically thick to thin transition\footnote{Our default assumption, based on previous work \citep[e.g.][]{russell2016}, is that this is synchrotron emission, but we discuss the possibility of coherent emission in the context of the giant flare below.}.
% ***We note that the radio flare peaks reported by \cite{russell2016} (between 4.6 GHz and 7.9 GHz) are all optically thin; they do not see any optically thick flares.
% SS Cyg subsequently went into radio quiescence or was highly self-absorbed at 15.5 GHz.
% The self-absorption scenario is possible since in the previous observations of SS Cyg, including VLBI, the radio core was detected through the end of the optical outburst \citep{russell2016}.
The multiple radio flares seen throughout the outburst are unlikely to be caused by the source crossing the ``jet line'' multiple times, given the evolution in the DFLD \citep{kording2008}, 
and may be either due to internal shocks, external shocks, or multiple discrete ejections not tied to any specific jet line.

In the first two hours of the AMI-LA observation on 29 February (MJD 57447.31), SS Cyg was at or below 150 $\mu$Jy.
% Say something about the big flare
On MJD 57447.4 we detected the onset of the giant flare (Figure~\ref{fig:bigflare_lc}).
After a gradual increase in flux density to 1.5 mJy, the giant flare reached its peak in $\sim$5 minutes (e-folding timescale of $\sim$1.5 minutes).
The implied variability brightness temperature is $T_B > 10^7$ K, which is comparable to or even greater than that seen in flares from magnetic CVs.
This value can also be compared with the lower limits to the brightness temperature placed from VLBA observations, $5.4\times 10^6$ K, and previously observed radio variability, $5.5\times10^3$ K \citep{russell2016}.
The $T_B$ of the giant flare and other radio flares detected during the outburst is consistent either with a synchrotron origin of the flares, as noted by previous works \citep{kording2008,russell2016}, 
or with a coherent source of emission.
Recently, 100\% circularly polarized radio flares were detected in nova-like cataclysmic variables \citep{coppejans2015}, indicating coherent emission.
We can neither confirm nor rule out the coherent mechanism hypothesis through our radio observations since the AMI-LA does not provide any polarization information.
% Our lower limit to $T_B$ does not rule out any contribution from coherent emission to the giant flare
The time resolution and sensitivity of the AMI-LA data are sufficient to (temporally) resolve the giant flare and get reliable spectral indices.
During the rise phase of the giant flare, the spectral index ($\alpha_{14.2GHz}^{16.8GHz}$) is $0.8\pm0.4$ (MJD 57447.4042; $S_{15.5GHz} = 6.9\pm0.3$ mJy), which increases to 
$2.8\pm0.3$ (MJD 57447.4047; $S_{15.5GHz} = 10.5\pm0.4$ mJy), consistent with a fully self-absorbed synchrotron source.
Beyond this maximum in the spectral index, the spectrum between 13--18 GHz evolves towards becoming optically thin. 
At the peak of the giant flare (MJD 57447.4064, $S_{15.5GHz} = 18.1\pm0.4$ mJy), the spectral index is $1.4\pm0.2$, and then decreases to a minimum of 
$-1.0\pm0.2$ (MJD 57447.4081; $S_{15.5GHz} = 13.7\pm0.4$ mJy) during the decay of the flare.
At MJD 57447.4103, another mJy-level flare, optically thin and peaking at $3.9\pm0.3$ mJy at 16.75 GHz, is detected and may have a delayed, less-luminous, counterpart at 14.25 GHz at 57447.4125 (see Figure~\ref{fig:bigflare_lc}).
Overall, the behavior of the radio emission at frequencies between 14.25 GHz and 16.75 GHz is consistent with the adiabatic expansion of a synchrotron-emitting plasma blob \citep{vanderLaan1966}.

% (***CAN WE DO ANY MODELING HERE?).
% Question: Can we disentangle some physics/energetics using the spectral indices?
% Question: Could the propeller model or magnetic reconnection be responsible for any of these observations?
% We note, however, that the spectral indices (Figure~\ref{fig:bigflare_lc}) suggest incoherent synchrotron as being the likely emission mechanism.

% X-ray, etc.
The question of why there was a giant radio flare towards the end of the outburst warrants some consideration.
According to the disk instability model\footnote{Although the DIM cannot explain anomalous outbursts very well \citep{schreiber2003}, 
the large flares occurs well after the slow-rise phase, once the outburst has apparently transitioned into a normal outburst.}, the material accreted from 
the companion star gradually builds up in the accretion disk until the temperature increases 
to the critical point to drive the hydrogen ionization instability, thereby transporting the accretion disk material rapidly onto the surface of the white dwarf.
During outburst, the X-ray emission rises at a similar time as the optical, and this is attributed to the material reaching the boundary layer.
The boundary layer becomes optically thick and quenches the X-ray radiation while the extreme UV radiation rises rapidly \citep{wheatley2003}.
Residual X-ray emission persists through the outburst phase.
Our X-ray observation was carried out soon after the optical light curve reached peak (MJD 57436.14).
The 1--10 keV luminosity, $1.7 \times 10^{31}$ erg s$^{-1}$, is in agreement with past observations during this phase of the outburst.
% Note that the implied X-ray-to-radio luminosity ratio of $10^{15.7}$ Hz (assuming a radio flux density of 200 $\mu$Jy) is in agreement coronal flares seen on \citep{benz1994}.
At the end of the optical and ultraviolet decay phases, the X-ray emission rises gradually, and later fades back to quiescent levels \citep{wheatley2003,russell2016}.
% Given the accretion rate prior to the decline phase of the outburst, the majority of the radio flares observed with the AMI-LA can be explained by jet activity.
During the phase when the optical and UV emission are declining, the accretion rate is expected to drop substantially.
Hence the ``giant'' flare, seen towards the end of the outburst, is intriguing.
If the giant flare is due to the ejection of plasma blob(s), then the large amplitude of the radio flare during this decay phase remains unexplained.
A comparison between our observations and a normal optical outburst suggests that the giant flare occurs either slightly before or right at the start of the 
X-ray increase at the end of the outburst \citep[see, e.g.][]{wheatley2003,russell2016}, 2--3 days before the peak of the X-ray emission.  
This suggests that the giant flare is not driven by the large change in the optical depth of the boundary layer.
Although the magnetic field of the white dwarf in SS Cyg is not high, it is possible that something like the propeller effect or magnetic reconnection takes over as the 
accretion rate is falling and generates an outflow of gas \citep[e.g. in cases of AE Aqr and VW Hyi;][]{wynn1997,warner2002,meintjes2000}, and hence a luminous flare in the radio.
%  (***CAN WE BE QUANTITATIVE HERE?)
% We note that there is also a hint (a few sigma) of a rise at the end of the VLA monitoring of the 2007 November outburst as well (K\"oring et al. in prep), and but it is after the 
Since the isotropic 15.5 GHz luminosity at the peak of the giant flare (and $\alpha_{14.2GHz}^{16.8GHz}\simeq1$) is $\sim10^{27}$ erg s$^{-1}$, this flare may be 
consistent with the L$_R$ to L$_X$ relationship \citep[using an estimate of the X-ray luminosity from][]{wheatley2003} seen for black hole systems \citep[see figure 9 of][]{russell2016}, albeit for a very short period of time.

% Overall interpretation

% This paragraph is reserved for the overall interpretation. The majority of the flares can be explained by DIM/accretion/jet, but what causes this ``giant'' flare? 
% Is there some material that comes pouring in from the inner accretion disk towards the end of the outburst? The giant flare may very well be a series of ejected blobs, not just one big blob.
% If we have no explanation, then we can simply say that this is an unexplained jet phenomenon.
% Question: Can we say something about the the ``boundary'' region with our data?

% compare with rusell2016 and kording2008
% - they see plateau (what is the signifcance; is it seen in all past observations?) - do we also see something similar?
% - timescales longer at lower frequencies? - consistent with jet

Lastly, we would like to highlight the importance of densely-sampled radio coverage, without which the rapid flares and the giant radio flare (which had a duration of 15 minutes) towards the end of 
SS Cyg's outburst would have remained unknown.
% Although we do not know whether there may have been another similar radio flare during the time when we were not observing, the probability is low, given our extensive coverage of the outburst.
The 200 hours of observing presented here may not be feasible with telescopes like the SKA or its pathfinders, and 
this underscores the need for interferometers like the AMI-LA to be operational in the era of large telescopes.
Niche areas of astronomy will continue to be accessible only with such small radio interferometers.

{\it Acknowledgements:
KPM's research is supported by the Oxford Centre for Astrophysical Surveys which is funded through the Hintze Family Charitable Foundation.
JCAMJ is the recipient of an Australian Research Council Future Fellowship (FT140101082).
RPF acknowledges support from the European Research Council Advanced Grant 267697 "4 Pi Sky: Extreme Astrophysics with Revolutionary Radio Telescopes".
TDR acknowledges support from the Netherlands Organisation for Scientific Research (NWO) Veni Fellowship, grant number 639.041.646.
AMS gratefully acknowledges support from the European Research Council under grant ERC-2012- StG-307215 LODESTONE. 
The AMI telescope is supported by the European Research Council under grant ERC-2012- StG-307215 LODESTONE, the UK Science and Technology Facilities Council (STFC) and the University of Cambridge.
% We acknowledge the use of optical light curves from the American Association of Variable Star Observers (AAVSO).
We extend special thanks to Stella Kafka for coordinating the optical observations on behalf of AAVSO, and to all the diligent AAVSO observers who contributed to the optical light curve.
We thank the AMI and Swift staff for scheduling the radio and X-ray observations.
We also thank the anonymous referee for provinding useful comments.}

\clearpage

\begin{table}
\centering
\caption{15.5 GHz AMI-LA measurements of the SS Cyg}
\label{tab:observations}
\begin{tabular}{lcrrrr}
\hline\hline  % Start and end MJD or middle MJD and duration might be better
MJD         & Dur.   & $S_{15.5GHz}$ & $\sigma_S$ & $\alpha_{14.2GHz}^{16.8GHz}$ & $\sigma_\alpha$ \\
            & (min)  & ($\mu$Jy)     &  ($\mu$Jy) &                              & \\
\hline
57420.5991 & 117 & 81 & 43 & \ldots & \ldots\\
57429.4993 & 598 & 57 & 17 & \ldots & \ldots\\
57430.5007 & 598 & 65 & 18 & \ldots & \ldots\\
57431.5124 & 596 & 41 & 22 & \ldots & \ldots\\
57432.4136 & 298 & 47 & 23 & \ldots & \ldots\\
57432.5690 & 149 & 90 & 31 & \ldots & \ldots\\
57432.6726 & 149 & 204 & 34 & \ldots & \ldots\\
57433.4925 & 598 & 69 & 18 & \ldots & \ldots\\
57434.3865 & 299 & 100 & 33 & \ldots & \ldots\\
57434.5423 & 149 & 174 & 36 & \ldots & \ldots\\
57434.6462 & 149 & 327 & 32 & 1.2 & 1.2 \\
57435.3398 & 149 & 191 & 30 & \ldots & \ldots\\
57435.4434 & 149 & 203 & 30 & \ldots & \ldots\\
57435.5210 & 74 & 420 & 42 & 1.3 & 0.9 \\
57435.5728 & 74 & 268 & 45 & \ldots & \ldots\\
57435.6505 & 149 & 171 & 50 & \ldots & \ldots\\
57436.4919 & 596 & 225 & 21 & \ldots & \ldots\\
57437.3896 & 298 & 154 & 29 & \ldots & \ldots\\
57437.5449 & 149 & 120 & 44 & \ldots & \ldots\\
57437.6226 & 74 & 305 & 54 & \ldots & \ldots\\
57437.6615 & 37 & 625 & 72 &  1.9 & 1.4 \\
57437.6874 & 37 & 222 & 74 & \ldots & \ldots\\
57439.3601 & 239 & 107 & 40 & \ldots & \ldots\\
57440.6077 & 317 & 117 & 38 & \ldots & \ldots\\
57441.4665 & 598 & 166 & 29 & \ldots & \ldots\\
57442.4577 & 580 & 192 & 23 & \ldots & \ldots\\
57443.4824 & 596 & 122 & 20 & \ldots & \ldots\\
57444.2729 & 37 & 451 & 77 & -2.5 & 1.7 \\
57444.2988 & 37 & 141 & 63 & \ldots & \ldots\\
57444.3376 & 74 & 126 & 45 & \ldots & \ldots\\
57444.4153 & 149 & 101 & 34 & \ldots & \ldots\\
57444.5707 & 298 & 149 & 24 & \ldots & \ldots\\
57445.2764 & 74 & 89 & 52 & \ldots & \ldots\\
57445.3284 & 74 & 448 & 40 & -0.5 & 0.9 \\
57445.3803 & 74 & 355 & 41 & -0.5 & 0.9 \\
57445.4323 & 74 & 112 & 43 & \ldots & \ldots\\
57445.5621 & 299 & 201 & 22 & \ldots & \ldots\\
57446.4529 & 598 & 70 & 17 & \ldots & \ldots\\
57447.3196 & 137 & 47 & 35 & \ldots & \ldots\\
57447.3793 & 34 & 147 & 67 & \ldots & \ldots\\
57447.3972 & 17 & 88 & 76 & \ldots & \ldots\\
57447.4046 & 4 & 9254 & 175 & \ldots & \ldots\\
57447.4076 & 4 & 15845 & 184 & \ldots & \ldots\\
57447.4106 & 4 & 2843 & 149 & \ldots & \ldots\\
57447.4136 & 4 & 983 & 123 & \ldots & \ldots\\
57447.4210 & 17 & 677 & 84 & \ldots & \ldots\\
57447.4285 & 4 & 1239 & 124 & \ldots & \ldots\\
57447.4315 & 4 & 433 & 137 & \ldots & \ldots\\
57447.4359 & 8 & 249 & 109 & \ldots & \ldots\\
57449.5854 & 266 & 91 & 96 & \ldots & \ldots\\
57450.4727 & 595 & 52 & 16 & \ldots & \ldots\\
57451.4792 & 536 & 2 & 18 & \ldots & \ldots\\
57452.4772 & 596 & 10 & 31 & \ldots & \\
\hline
\multicolumn{6}{p{2.8in}}{Notes: a) $S_{15.5GHz}$ is the peak pixel values at the location of SS Cyg. Flux density values that are $<$2$\sigma$ are to be considered as non-detections. b) $\sigma_S$ is the RMS noise. 
c) The spectral index values have large uncertainties except in the cases of flare peaks, and only those values are noted here.}
\end{tabular}
\end{table}

\begin{table}
\centering
\caption{15.5 GHz finely sampled measurements of the ``giant'' flare}
\label{tab:observations_giantflare}
\begin{tabular}{lcrrrr}
\hline\hline  % Start and end MJD or middle MJD and duration might be better
MJD         & Dur.   & $S_{15.5GHz}$ & $\sigma_S$ & $\alpha_{14.2GHz}^{16.8GHz}$ & $\sigma_\alpha$ \\
            & (min)  & ($\mu$Jy)     &  ($\mu$Jy) &                              & \\
\hline
\multicolumn{6}{c}{}\\
\hline
57447.39745 & 0.8 & 547 & 290 & \ldots & \ldots\\
57447.39800 & 0.8 & 312 & 291 & \ldots & \ldots\\
57447.39856 & 0.8 & 252 & 281 & \ldots & \ldots\\
57447.39912 & 0.8 & 531 & 295 & \ldots & \ldots\\
57447.39968 & 0.8 & 285 & 280 & \ldots & \ldots\\
57447.40024 & 0.8 & 450 & 392 & \ldots & \ldots\\
57447.40192 & 0.8 & 703 & 361 & \ldots & \ldots\\
57447.40248 & 0.8 & 1792 & 315 & \ldots & \ldots\\
57447.40304 & 0.8 & 2010 & 346 & \ldots & \ldots\\
57447.40359 & 0.8 & 4090 & 335 & 0.8 & 0.9\\
57447.40415 & 0.8 & 6860 & 326 & 0.8 & 0.5\\
57447.40471 & 0.8 & 10533 & 393 & 2.8 & 0.3\\
57447.40527 & 0.8 & 13288 & 368 & 2.0 & 0.2\\
57447.40583 & 0.8 & 15963 & 397 & 1.2 & 0.2\\
57447.40639 & 0.8 & 18081 & 419 & 1.4 & 0.2\\
57447.40695 & 0.8 & 16563 & 378 & 0.5 & 0.2\\
57447.40751 & 0.8 & 16816 & 404 & 0.9 & 0.2\\
57447.40807 & 0.8 & 13745 & 355 & -1.0 & 0.2\\
57447.40862 & 0.8 & 12180 & 869 & -1.2 & 0.8\\
57447.40974 & 0.8 & 3272 & 581 & \ldots & \ldots\\
57447.41030 & 0.8 & 4674 & 327 & 0.9 & 0.8\\
57447.41086 & 0.8 & 3462 & 300 & 1.5 & 1.1\\
57447.41142 & 0.8 & 1859 & 335 & \ldots & \ldots\\
57447.41198 & 0.8 & 2085 & 344 & \ldots & \ldots\\
57447.41254 & 0.8 & 1719 & 306 & \ldots & \ldots\\
57447.41310 & 0.8 & 1699 & 326 & \ldots & \ldots\\
57447.41366 & 0.8 & 1220 & 279 & \ldots & \ldots\\
57447.41421 & 0.8 & 1355 & 327 & \ldots & \ldots\\
57447.41477 & 0.8 & 1337 & 292 & \ldots & \ldots\\
57447.41590 & 2.4 & 1117 & 199 & \ldots & \ldots\\
57447.41758 & 2.4 & 1399 & 347 & \ldots & \ldots\\
57447.41926 & 2.4 & 677 & 176 & \ldots & \ldots\\
57447.42094 & 2.4 & 806 & 167 & \ldots & \ldots\\
57447.42261 & 2.4 & 1225 & 180 & \ldots & \ldots\\
57447.42429 & 2.4 & 1170 & 205 & \ldots & \ldots\\
57447.42597 & 2.4 & 544 & 252 & \ldots & \ldots\\
57447.42765 & 2.4 & 1409 & 187 & \ldots & \ldots\\
57447.42932 & 2.4 & 1155 & 197 & -0.9 & 1.7\\
57447.43100 & 2.4 & 1079 & 188 & \ldots & \ldots\\
57447.43268 & 2.4 & 763 & 258 & \ldots & \ldots\\
57447.43436 & 2.4 & 1201 & 253 & \ldots & \ldots\\
57447.43604 & 2.4 & 645 & 188 & \ldots & \ldots\\
57447.43772 & 2.4 & 614 & 202 & \ldots & \ldots\\
57447.43939 & 2.4 & 1154 & 229 & \ldots & \ldots\\
57447.44107 & 2.4 & 275 & 373 & \ldots & \ldots\\
57447.44275 & 2.4 & 558 & 260 & \ldots & \ldots\\
57447.44443 & 2.4 & 457 & 214 & \ldots & \ldots\\
57447.44611 & 2.4 & 400 & 195 & \ldots & \ldots\\
57447.44778 & 2.4 & 352 & 212 & \ldots & \ldots\\
57447.44946 & 2.4 & 195 & 468 & \ldots & \ldots\\
57447.45114 & 2.4 & 425 & 175 & \ldots & \ldots\\
57447.45282 & 2.4 & 384 & 192 & \ldots & \ldots\\
57447.45450 & 2.4 & 757 & 201 & \ldots & \ldots\\
57447.45617 & 2.4 & 333 & 178 & \ldots & \ldots\\
57447.45785 & 2.4 & 770 & 520 & \ldots & \ldots\\
57447.45953 & 2.4 & 142 & 184 & \ldots & \ldots\\
57447.46121 & 2.4 & 420 & 210 & \ldots & \ldots\\
57447.46289 & 2.4 & 25 & 173 & \ldots & \ldots\\
57447.46456 & 2.4 & 420 & 192 & \ldots & \ldots\\
57447.46624 & 2.4 & 241 & 279 & \ldots & \ldots\\
57447.46792 & 2.4 & 89 & 148 & \ldots & \ldots\\
\hline
% \multicolumn{5}{p{2.8in}}{Notes:}
\multicolumn{6}{p{2.8in}}{Notes: a) $S_{15.5GHz}$ is the peak pixel values at the location of SS Cyg. b) $\sigma_S$ is the RMS noise. c) Only the spectral index values having uncertainties less than two are noted here.}
\end{tabular}
\end{table}

\end{document}